\documentclass[twocolumn,showpacs,preprintnumbers,amsmath,amssymb]{revtex4}

\usepackage{graphicx}% Include figure files
\usepackage{dcolumn}% Align table columns on decimal point
\usepackage{bm}% bold math

\begin{document}

\preprint{}

\title{An on-chip coupled resonator optical waveguide single-photon buffer}

\author{Hiroki Takesue$^{1}$} \email{takesue.hiroki@lab.ntt.co.jp}
\author{Nobuyuki Matsuda$^{1,2}$}
\author{Eiichi Kuramochi$^{1,2}$}
\author{William J. Munro$^{1}$}
\author{Masaya Notomi$^{1,2}$}

\affiliation{%
$^1$NTT Basic Research Laboratories, NTT Corporation, \\
3-1 Morinosato Wakamiya, Atsugi, Kanagawa, 243-0198, Japan\\
$^2$NTT Nanophotonics Center, NTT Corporation, \\
3-1 Morinosato Wakamiya, Atsugi, Kanagawa, 243-0198, Japan
}

\begin{abstract}
Integrated quantum optical circuits are now seen as one of the most promising approaches with which to realize single photon quantum information processing. Many of the core elements for such circuits have been realized including sources, gates and detectors. However, a significant missing function necessary for photonic information processing on-chip is a buffer, where single photons are stored for a short period of time to facilitate circuit synchronization. Here we report an on-chip single photon buffer based on coupled resonator optical waveguides (CROW) consisting of 400 high-Q photonic crystal line defect nanocavities. By using the CROW, a pulsed single photon was successfully buffered for 150 ps with 50-ps tunability while maintaining its non-classical properties. Furthermore, we showed that our buffer preserves entanglement by storing and retrieving one photon from a time-bin entangled state. This is a significant step towards an all-optical integrated quantum information processor.
\end{abstract}

\maketitle

%\section*{Introduction}

\begin{figure*}[thb]

\centerline{\includegraphics[width=.9\linewidth]{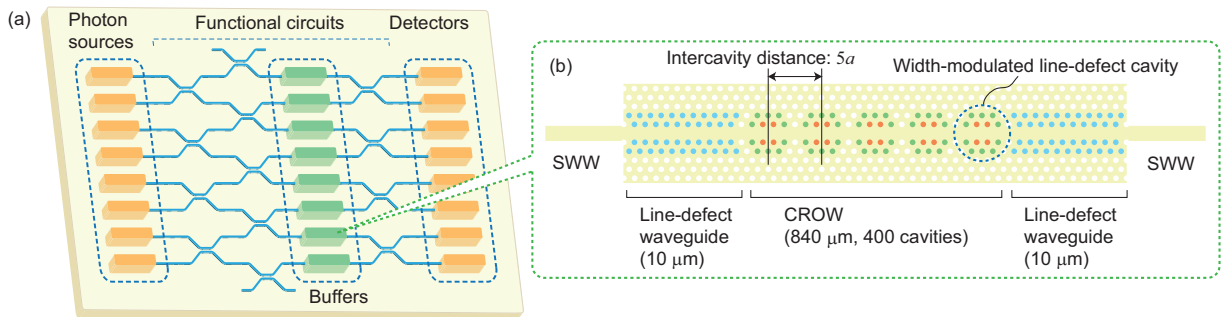}}

\caption{(a) A conceptual diagram of a photonic quantum processor based on integrated quantum photonics including the use of quantum buffers. In principle such buffers could be placed at any point in the circuit. (b) Quantum buffer based on CROW, which consists of photonic crystal nanocavities. Silicon wire waveguides (SWW) are integrated with the PhC section to access the CROW. 
%$N$: number of cavities, $L_{cc}$: intercavity distance. 
}
\label{1}

\end{figure*}

Photonic quantum information processing is now recognized as one of the best ways to achieve large-scale quantum computation \cite{kok} and communication \cite{gisin}. 
To date, most photonic quantum information experiments have been performed using bulk optics \cite{kok,pan}. As the size and complexity of these realizations increase, these bulk optics approaches begin to limit and hamper the functionally of the experiments that can be performed. 
The natural solution is to move to an integrated photonics approach \cite{plc}.
Such approaches have already shown that on-chip photon sources \cite{matsuda,orieux,valles}, detectors \cite{sspd,sspd2} and circuit-based gates can be achieved. Non-trivial circuits have been implemented ranging from basic observations of quantum interference \cite{plc,multi} and multi-photon entanglement manipulation \cite{matt}, 
to the sophisticated implementation of quantum computation tasks such as Shor's algorithm \cite{politi}, quantum walks \cite{peruzzo,crespi}, and boson sampling \cite{spring,tillmann}. 
There is, however, a critical missing element that has yet to be realized on-chip, that is a quantum buffer. 
Figure \ref{1} (a) is a schematic of an all-on-chip quantum processor expected in the near future, where quantum functional circuits are integrated with quantum buffers together with photon sources and detectors. 
A buffer is critical for synchronizing the photons in such an integrated circuit, which becomes much more difficult as the integration and compactness of circuit increases. 
The buffer can also be used to unsynchronize (temporally stagger) photons so they do not interfere in certain parts of the circuits.
Furthermore, a tunable buffer together with other active elements such as optical switches \cite{shadbolt} enables us to realize a ``quantum FPGA", namely a fully programmable integrated quantum optical circuit.

A natural way to realize an integrated single photon buffer is to use the slow-light effect in optical waveguides \cite{baba}. 
Among several waveguide-based slow light devices reported so far \cite{notomidisp,ynu1}, a coupled resonator optical waveguide (CROW) is a promising candidate due to its large bandwidth with small group velocity dispersion \cite{yariv,karle,hara,xia,notomi,cooper}. A CROW is a one-dimensional array of identical optical cavities, where adjacent cavities are coupled to each other. 
An extended mode along the waveguide can be formed using a nearest-neighbour interaction. 
Consequently, a CROW exhibits a transmission bandwidth that is far larger than the individual cavities, while the group velocity is significantly reduced inside the band \cite{yariv}.

In this paper, we will demonstrate an integrated single photon buffer based on a CROW fabricated using silicon photonic crystal (PhC) technologies \cite{notomi}. With this CROW, whose length was only 840 $\mu$m, a pulsed photon from a correlated photon pair source was buffered by up to 150 ps with a tunability of 50 ps, while preserving its quantum correlation with the other photon. Furthermore, we confirmed that entanglement could be preserved after storing and retrieving one photon from a time-bin entangled state. 

%%%%%%%%%%%%%%%%%%%%%%%%%%%%%%%%%%%%%

\section*{Results}

\subsection*{CROW based on silicon photonic crystal waveguide}

Figure \ref{1} (b) shows the design of our CROW \cite{notomi}, which is based on a width-modulated line-defect cavity in a silicon PhC with a two-dimensional triangular-lattice of air holes \cite{tanabe}. The lattice constant $a$ of our PhC waveguide is 420 nm, and the intercavity distance is $5a$. The Q of each cavity is $\sim 10^6$. The number of cavities is 400, which means that the total length is 840 $\mu$m. A 10-$\mu$m PhC line defect waveguide is fabricated at each end of the CROW. 
%the following part is moved from Method
A transmission band was formed between 1543 and 1548 nm. Although a peaky structure that was probably caused by fabrication errors was observed \cite{notomi}, there were several low-loss peaks in the spectrum. 
In the following experiments, we tuned the chip temperature so that one of the spectral peaks matched the signal photon wavelength of 1546.70 nm. 
The average loss of the peaks was approximately 26 dB, which includes an $\sim 8$-dB coupling loss between the lensed fibres and the waveguide at each facet. 
The observed temperature dependence of the transmission spectrum was $\sim 0.07$ nm/$^\circ$C. 
The PhC section is integrated with silicon wire waveguides (SWW). 
We fabricated another waveguide of the same length on the same chip as a reference for the temporal delay, where we replaced the CROW section with a PhC line defect waveguide. 
%(The CROW characteristics are detailed in Methods.) 

\begin{figure}[thb]

\centerline{\includegraphics[width=\linewidth]{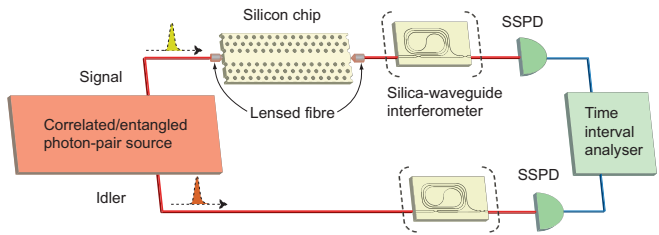}}

\caption{Experimental set-up for observing photon buffering and entanglement storage. The silica-waveguide interferometers were inserted in the entanglement storage experiment. }
\label{2}

\end{figure}

\subsection*{Observation of photon buffering}

\begin{figure*}[thb]

\centerline{\includegraphics[width=.9\linewidth]{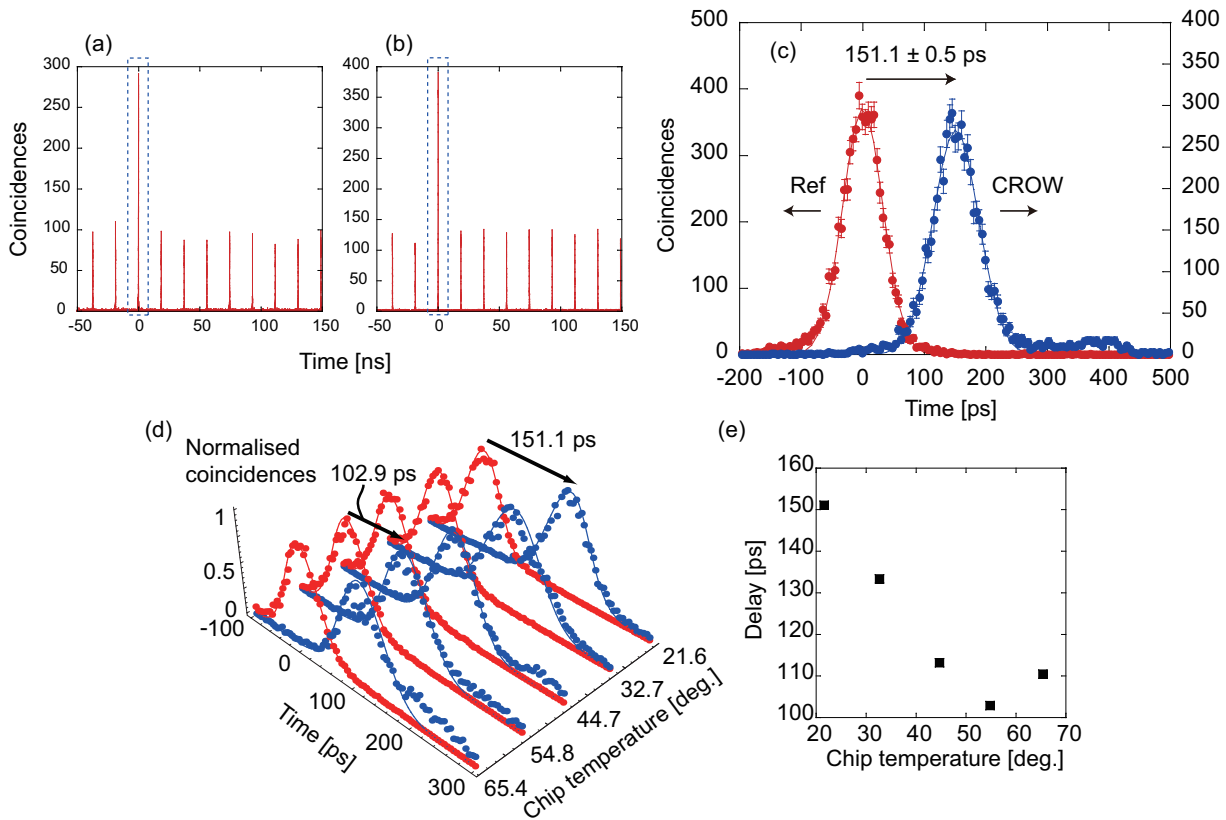}}

\caption{(a) and (b) Time interval histograms when the signal photons were transmitted through (a) the CROW and (b) the reference waveguide. (c) Enlarged image of main peaks. The blue and red dots show histograms obtained with signal photons transmitted through the CROW and the reference waveguide, respectively. Statistical error bars are shown, and the lines indicate Gaussian fitting to the data. (d) Time interval histograms around the main peaks with various chip temperatures. While the temporal positions of the peaks for the reference waveguide (red dots) remained unchanged, those for the CROW (blue dots) showed strong temperature dependence. The delay as a function of chip temperature is plotted in (e). 
%The maximum and minimum delay times were 151.1 (21.6$^\circ$C) and 102.9 (54.8$^\circ$C), respectively. 
For figures (a)-(d), coincidences were counted for 1 million start signals. }
\label{3}
\end{figure*}

Our experimental set-up for the observation of photon buffering is depicted in Fig. \ref{2}. A photon-pair source based on spontaneous four-wave mixing (SFWM) in a dispersion shifted fibre (DSF) (see Methods for details) generates signal and idler photons. 
The wavelengths of the signal and idler photons were 1546.70 and 1555.53 nm, respectively, with a bandwidth of 0.2 nm. The signal photon was coupled to the silicon chip that included the CROW and the reference waveguide by using lensed fibres, and the photons output from the waveguide were received by a superconducting single photon detector (SSPD) \cite{golt}. The idler photon was directly measured by the second SSPD. 
The detection signal from the SSPD for the signal (idler) was used as the start (stop) signal for a time interval analyser performing coincidence measurements.

Figure \ref{3} shows time interval histograms of the coincidence counts as a function of the relative delay between the signal and idler detection events. The average signal photon number per pulse was set at 0.13. Figure \ref{3} (a) and (b) show the time interval histograms where the signal photons were transmitted through the CROW and the reference waveguide, respectively. The chip temperature was 21.6$^\circ$C. 
We observed higher peaks at a relative delay near zero (highlighted by dotted boxes), which correspond to the coincidences caused by the correlated signal and idler photons generated by the same pump pulse. These peaks are significantly higher than other peaks that correspond to accidental coincidences caused by photons generated by different pump pulses, implying the existence of an intensity correlation between the two photons. The cross correlation $g_{si}^{(2)} (0)=P_{si}/P_s P_i$, where $P_{si}$ is the coincidence detection probability and $P_s$ $(P_i)$ is the probability of detecting a signal (idler) photon, was measured and found to be $3.25 \pm 0.06$ for the CROW and $3.10 \pm 0.05$ for the reference waveguide. When we removed the waveguide from the set-up, $g_{si}^{(2)}(0)$ was $3.22 \pm 0.05$. Note that $g_{si}^{(2)} (0)\geq 2$ implies the existence of a non-classical intensity correlation between the signal and idler photons. Thus, our result confirms that the quantum correlation was preserved even after the signal photon had been transmitted through 400 high-Q nanocavities, with virtually no degradation.

The coincidences at the main peaks were regarded as the detection events of quasi-single photons heralded by the detection of the idler photon. 
Since the detection times of the idler photons were unchanged, we could measure the delay time of the quasi-single photons in the CROW from the temporal shift of the main peaks. Figure \ref{3} (c) shows an enlarged image of the main peaks in Fig. \ref{3} (a) and (b), where a clear separation of the two peaks is seen. This result confirmed that the CROW, whose physical length was only 840 $\mu$m, stored a single photon for as long as $151.1 \pm 0.5$ ps. 
The group index of the reference waveguide was approximately 5, which means that the speed of the pulsed photon in the CROW was slowed to 1/59 of the light speed in a vacuum. %we obtained an effective group index for the CROW of 59. 
The $1/e$ half widths of the peaks were $52.7 \pm 0.6$ ps (CROW) and $46.3 \pm 0.4$ (reference waveguide). We also observed a slight tail in the coincidence histogram for the CROW. These observations suggest that the small spectral structure of the CROW may have led to the excess broadening of the temporal width and the slight distortion of the pulse shape. 
Still, the obtained result indicates that our buffer can store a broadband single photon pulse whose $1/e$ half temporal width is smaller than $< 30$ ps (see Methods).

%%%%%%%%%%%%%%%%%%%%%%%%%%%%%%%%%%%%%%%%%%%%%%%%%%%%%%%%%%
\subsection*{Delay tuning}

We now demonstrate the tunability of our single photon buffer. The transmission band of our silicon PhC CROW shifts to a longer wavelength as the chip temperature changes, and the wavelength dependence of the group index forms a sinusoidal curve with its minimum value at the centre of the band \cite{yariv}. This implies that we can shift the dispersion characteristic by changing the temperature, which enables us to tune the buffer time. 
We varied the chip temperature from 21.6 to 65.4$^\circ$C, while maintaining the wavelength of the signal photon. The histograms obtained around the main peaks are shown in Fig. \ref{3} (d), which clearly reveals a temperature-dependent delay. We also confirmed the preservation of a non-classical correlation with the idler photons for all chip temperatures. 
As shown in Fig. \ref{3} (e), the buffer time was varied from 103 to 151 ps, suggesting that we realized a tunable buffer with a tuning range of approximately 50 ps. In free space, such a 50-ps tunability requires a 1.5-cm variable delay compared with 840 $\mu$m in our waveguide.

%%%%%%%%%%%%%%%%%%%%%%%%%%%%%%%%%%%%%%%%%%%%%%%%%%%%%%%%%%%%%%%%%%%%%%%%%%
\subsection*{Storage of entanglement}

We have already demonstrated the tunable buffering ability of our CROW. However, if they are to be used in photonic circuits for quantum information tasks, photons buffered by the CROW must preserve quantum coherence with other photons in the circuit. The simplest demonstration of this is to buffer one half of an entangled state and then show that the buffered photon is still entangled after it is released from the CROW. 
Again using SFWM in a DSF, we generated a high-dimensional time-bin entangled state at a clock frequency of 2 GHz, whose state is given by \cite{sequential} (see Methods),  
\begin{equation}
|\Psi\rangle = \frac{1}{\sqrt{M}} \sum_{k=1}^M |k\rangle_s |k\rangle_i. \label{ent}
\end{equation}
Here, $M$ is the number of time slots where the coherence of the pump pulse is preserved and $|k\rangle_x$ denotes a state where there is a photon at the $k$th time slot in the mode $x$ ($s$: signal, $i$: idler).
%The signal photon was coupled to the silicon chip, whose temperature was set at 21.6$^\circ$C. 
As shown in Fig. \ref{2}, the signal photons transmitted through the silicon chip (temperature: 21.6$^\circ$C) and idler photons were input into delayed Mach-Zehnder interferometers whose propagation time difference between two arms was 1 ns. 
These interferometers were made of silica-based optical waveguides fabricated on silicon substrates, and the phase difference between two arms was tuned by changing the waveguide temperature \cite{timebin}. 
The photons from the interferometers were detected by SSPDs, and the coincidences were counted. 
The coincidence probability $P$ is given by (see Methods for details)
\begin{equation}
P \propto 1 + V \cos (\phi_s + \phi_i), \label{coin}
\end{equation}
where $V$ is the two-photon interference visibility, and $\phi_s$ and $\phi_i$ denote the phase difference between the two arms of the interferometers for the signal and the idler, respectively.

\begin{figure}[thb]

\centerline{\includegraphics[width=.8\linewidth]{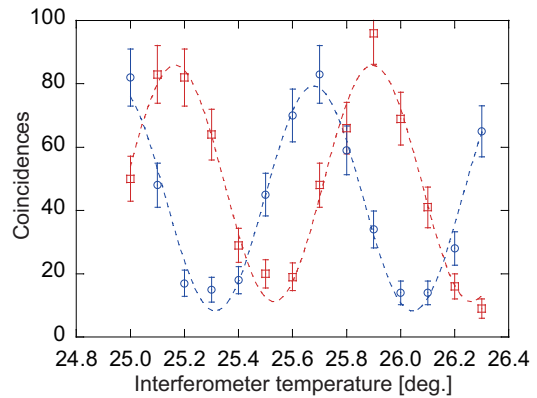}}

\caption{Two-photon interference fringes. Squares and circles show the results when the idler interferometer temperature was tuned to 22.74 and 22.94$^\circ$C, respectively. The visibilities of fitted curves are $77 \pm 5$\% (22.74$^\circ$C) and $81 \pm 5$\% (22.94$^\circ$C), which indicates that that the stored entanglement should violate Bell's inequality. The coincidences were counted for 500,000 start signals. }
\label{4}
\end{figure}

We measured the coincidences as a function of the signal interferometer temperatures while fixing the idler interferometer temperature at 22.74 and 22.94$^\circ$C, which correspond to two non-orthogonal measurements for the idler photons. The two-photon interference measurement is described in detail in Methods. The result is shown by squares (22.74$^\circ$C) and circles (22.94$^\circ$C) in Fig. \ref{4}, where clear modulations of the coincidence counts can be observed. The data are fitted with Eq. (\ref{coin}), and the visibilities of the fitted curves were $77 \pm 5$\% (22.74$^\circ$C) and $81 \pm 5$\% (22.94$^\circ$C). 
These visibilities suggest that the buffer could preserve high-quality entanglement that violates the Bell's inequality. 

\section*{Discussion}

%extension of delay time
It is possible to increase the delay time further. One way to achieve this is to increase the length of the CROW at the expense of an increased device footprint. Another way is to increase the group index by decreasing the cavity coupling \cite{yariv}, which can be realized by increasing the intercavity distance \cite{notomi}. In this case, we can retain the small device footprint, but the transmission bandwidth is generally reduced. Thus, we can design a CROW-based single-photon buffer flexibly depending on the requirements of a particular quantum information system. 

Although the temperature tuning technique demonstrated here is relatively slow (tuning time $\sim$ 1 ms), it is sufficient for realizing programmable integrated quantum optical circuits. Nevertheless, a buffer with a faster tuning speed will find more applications for example an active quantum feedforward \cite{feedforward} on a chip. A possible candidate for a fast tunable buffer is a high-Q PhC cavity dynamically tuned by carrier injection \cite{tanabe}.

In conclusion, we have demonstrated a single photon buffer based on a CROW. We showed that a pulsed single photon was buffered for up to 150 ps with 50-ps tunability, and confirmed the preservation of entanglement. Also note that the present experiment is the first demonstration of non-classical light propagation in a slow-light waveguide. We believe that such a quantum buffer will be useful for realizing a flexible and reconfigurable quantum processor on a chip.

%possibility of on-demand buffer

\newpage
\section*{Methods}

\noindent
%{\bf Details of CROW.} 
%A transmission band was formed between 1543 and 1548 nm. Although a peaky structure that was probably caused by fabrication errors was observed \cite{notomi}, there were several low-loss peaks in the spectrum. 
%In the above experiments, we tuned the chip temperature so that one of the spectral peaks matched the signal photon wavelength of 1546.70 nm. 
%The average loss of the peaks was approximately 26 dB, which includes an $\sim 8$-dB coupling loss between the lensed fibres and the waveguide at each facet. 
%The observed temperature dependence of the transmission spectrum was $\sim 0.07$ nm/$^\circ$C. 
%We fabricated another waveguide of the same length on the same chip as a reference for the temporal delay, where we replaced the CROW section with a PhC line defect waveguide. 

\vspace{3mm}

\noindent
{\bf Generation of correlated/entangled photon pairs using dispersion shifted fibre.} 
A pump pulse train with a wavelength of 1551.1 nm was amplified by an erbium-doped fibre amplifier (EDFA), and then transmitted through optical filters to suppress the amplified spontaneous emission noise generated in the EDFA. The filtered pulse train was launched into a 500-m-long DSF. Through the SFWM process in the DSF, two pump photons were annihilated and a correlated photon pair was generated. The generated photons were input into another filter to separate the signal and idler photons, whose wavelengths were 1546.70 and 1555.53 nm, respectively, and the spectral width was 0.2 nm for both channels. The signal and idler photons were detected by respective SSPDs, whose detection efficiencies were 14 and 11\% for the signal and idler channels, respectively. The dark count rate was $\sim$10 cps for both detectors. 

We used different pump pulse sources in photon buffer experiments and entanglement storage experiments. 
In the photon buffer experiments described above, we used sub-ps pulses with a repetition frequency of 53.65 MHz generated from a fibre mode-locked laser. The pulses were transmitted through an optical filter whose spectral width was 0.2 nm, before being input into the EDFA. The full width at half maximum (FWHM) of the pump pulse was measured to be $\sim$ 20 ps with an autocorrelator. In the entanglement storage experiments, we used continuous-wave laser light from an external-cavity diode laser, which was modulated into 60-ps, 2-GHz frequency pulses using a lithium niobate intensity modulator. 

In the entanglement storage experiment, the average signal photon number per pulse was set at 0.01, and the chip temperature was 21.6$^\circ$C. The pump coherence time was $\sim 10$ $\mu$s, which means that $M$ in Eq. (\ref{ent}) was $\sim$ $2 \times 10^4$. We used silica-waveguide interferometers with a 1-ns propagation time difference between two arms, while the temporal interval of our entangled photon pulses was 0.5 ns. This means that a state $|k\rangle_x$ ($x=s$: signal, $i$: idler) is converted to $(|k\rangle_x + e^{i \phi_x} |k+2\rangle_x)/2$, where $\phi_x$ is the phase difference between the two arms of the interferometer. Then, Eq. (\ref{ent}) is converted to 
\begin{equation}
|\Psi\rangle \to |1\rangle_s |1\rangle_i + \sum_k^{M+1} (1+e^{i(\phi_s + \phi_i)}) |k\rangle_s |k\rangle_i + |M+2\rangle_s |M+2\rangle_i,
\end{equation}
where only the terms that contributed to the coincidence detection are shown and the normalizing term is discarded for simplicity. 
By ignoring the first and the last terms of the right hand side of the above equation, we obtain the coincidence probability $P$ as Eq. (\ref{coin}). 
%\begin{equation}
%P \propto 1 + V \cos (\phi_s + \phi_i), \label{coin}
%\end{equation}
%where $V$ is the two-photon interference visibility. 
We employed the high-dimensional time-bin entangled state because we can fully use the time domain and thus achieve fast measurement despite the relatively small average photon number per pulse.

\noindent
{\bf Estimation of photon pulse broadening in CROW.} 
The $1/e$ half width of the SSPD jitter, $\sigma_{sspd}$, was measured and found to be $\sim$30 ps for both channels. Assuming Gaussian jitter and photon pulse broadening, the $1/e$ half widths of the main peaks of time interval histograms, $\sigma_{main}$, can be expressed as
\begin{equation}
\sigma_{main}^2 = \sigma_s^2 + \sigma_i^2 + 2 \sigma^2_{sspd},
\end{equation}
where $\sigma_s$, $\sigma_i$ denote the $1/e$ half widths of the signal and idler photons, respectively. We assume that the idler photon width was the same as the pump pulse width (20 ps FWHM), which means $\sigma_i \simeq 12$ ps. Thus we obtain $\sigma_s \simeq 28$ ps for the signal photon width after transmission in the CROW.

%\clearpage

\section*{Acknowledgements}

H. T. thanks Drs. H. Sumikura and T. Inagaki for their help in SSPD operation. 
This work was supported by the Japan Society for the Promotion of Science with a Grant-in-Aid for Scientific Research (No. 22360034).

%\section*{Competing financial interests}

%The authors declare that there are no competing financial interests.

%\section*{Supplementary Information}

\end{document}